\documentclass{agujournal}

\journalname{JGR}

\usepackage{hyperref}
\usepackage{textcomp}
\usepackage{graphicx}
\usepackage{color}

\begin{document}

\title{Three-Dimensional X-line Spreading in Asymmetric Magnetic Reconnection}
\authors{Tak~Chu~Li\affil{1}, Yi-Hsin Liu\affil{1}, Michael Hesse\affil{2,3}, Ying Zou\affil{4,5,6}}

\affiliation{1}{Dartmouth College, Hanover, New Hampshire, USA}
\affiliation{2}{University of Bergen, Bergen, Norway}
\affiliation{3}{Southwest Research Institute, San Antonio, Texas, USA}
\affiliation{4}{Boston University, Boston, Massachusetts, USA}
\affiliation{5}{University Corporation for Atmospheric Research, Boulder, Colorado, USA}
\affiliation{6}{University of Alabama in Huntsville, Huntsville, Alabama, USA}

\correspondingauthor{Tak Chu Li}{tak.chu.li@dartmouth.edu}

\begin{keypoints}
\item The continuous onset of reconnection is found to be important for 3D X-line spreading in large-scale particle-in-cell simulations.
\item Simulations with fast (slow) onset demonstrate X-line spreading at the Alfv\'en speed based on the guide field $V_{Ag}$ (sub-Alfv\'enic, ion/electron drift speeds).
\item The current sheet thickness and collisionless tearing growth rate play a key role in determining the spreading speeds.
\end{keypoints}

\begin{abstract}
The spreading of the X-line out of the reconnection plane under a strong guide field is investigated using large-scale three-dimensional (3D) particle-in-cell (PIC) simulations in asymmetric magnetic reconnection. A simulation with a thick, ion-scale equilibrium current sheet (CS) reveals that the X-line spreads at the ambient ion/electron drift speeds, significantly slower than the Alfv\'en speed based on the guide field $V_{Ag}$. Additional simulations with a thinner, sub-ion-scale CS show that the X-line spreads at $V_{Ag}$ (Alfv\'enic spreading), much higher than the ambient species drifts. 
An Alfv\'enic signal consistent with kinetic Alfv\'en waves develops and propagates, leading to CS thinning and extending, which then results in reconnection onset. The continuous onset of reconnection in the signal propagation direction manifests as Alfv\'enic X-line spreading.
The strong dependence on the CS thickness of the spreading speeds, and the X-line orientation are consistent with the collisionless tearing instability. Our simulations indicate that when the collisionless tearing growth is sufficiently strong in a thinner CS such that $\gamma/\Omega_{ci}\gtrsim\mathcal{O}(1)$, Alfv\'enic X-line spreading can take place. Our results compare favorably with a number of numerical simulations and recent magnetopause observations. A key implications is that the magnetopause CS is typically too thick for Alfv\'enic X-line spreading to effectively take place.
\end{abstract}


\section{Introduction}\label{intro}
Magnetic reconnection is an important process of converting magnetic energy into particle bulk flow and thermal energies. It is a primary driver of space weather surrounding the Earth. While two-dimensional (2D) models have been widely used to describe the essential aspects of reconnection \citep{birn01a}, the three-dimensional (3D) dynamics represents a frontier of current reconnection research (e.g.,\citep{daughton11a,yhliu13a,YHLiu:19a,TKMNakamura:2016,price16a,Dahlin:16,Janvier:2017}). In this work, we focus on the evolution of the extension of a localized reconnection X-line in the out-of-plane direction. This dynamic process is 3D \emph{X-line spreading}. 3D X-line spreading is important for the coupling between global dynamics and local kinetic physics of reconnection.

3D X-line spreading is observed in a variety of contexts. On the sun, solar flare ribbons are observed to spread along the magnetic polarity inversion line unidirectionally or bidirectionally \citep{Isobe:02b,Cheng:12,Fletcher:04,LeeGary:08,LiuC:10,Qiu:09,Qiu:17}. In the Earth's magnetotail and magnetopause, a wide range of X-line extents ranging from a few Earth radii $R_E$ to longer than 10$R_E$ along the current (out-of-plane) direction exists \citep{Phan:00,Fuselier:02,Nakamura:04,Li:13a,Zou:2019}. In the solar wind, a long extended X-line of over 100's of $R_E$ has also been reported \citep{Phan:06}. In the laboratory, experiments such as the Magnetic Reconnection Experiment (MRX) \citep{Dorfman:13,Dorfman:14} and the Versatile Toroidal Facility (VTF) \citep{Katz:10,Egedal:11} have also observed X-line spreading across the device.

Numerical simulations have been used to investigate 3D X-line spreading. 3D fully kinetic simulations have reported the tendency of merging of multiple patchy reconnection sites \citep{Hesse:01b,Hesse:05}, and the tendency of extending of locally driven reconnection regions \citep{Pritchett:01b} along the electron current direction. Hall-MHD simulations also show that a localized X-line propagates as a wave structure in the electron current direction \citep{Huba:02,Huba:03}. It is later confirmed that the X-line can extend in both the ion and electron current directions in 3D hybrid simulations \citep{Karimabadi:04} and two-fluid simulations \citep{shay03a}. The former further shows that the X-line only extends to where resistivity is present. When a thicker current sheet is initialized, significant slowdown of the X-line spreading is observed in 3D PIC simulations \citep{Lapenta:06}. Recent systematic study using Hall-MHD simulations shows that the X-line spreads at the ion/electron flow speeds \citep{TKMNakamura12}, establishing a quantitative measure of the spreading speeds. While the above mentioned studies have used zero or moderate guide field, intended for magnetotail applications, X-line spreading under a strong guide field has been recently studied. Two-fluid simulations using varying guide fields found that the spreading speed of the X-line scales approximately linearly with the guide field strength; it is suggested that the X-line spreads at the faster of the guide-field Alfv\'en speed and the ion/electron flow speed \citep{Shepherd:12}.

Recent magnetopause observations using THEMIS and SuperDARN radars, however, found that the spreading speed is significantly lower than the guide-field Alfv\'en speed but consistent with the ion/electron drift speed \citep{Zou:2018}. On the sun, the spreading of solar flare ribbons is reported to have a maximum speed of nearly an order of magnitude smaller than the characteristic coronal Alfv\'en speed \citep{Qiu:17}. On the other hand, in laboratory experiments, it has been observed that reconnection under strong guide field conditions spreads at approximately twice the guide-field Alfv\'en speed \citep{Katz:10}. How does the X-line spread under a strong guide field and under what conditions does it spread at the guide-field Alfv\'en speed? This work aims to address them using large-scale 3D PIC simulations with varying CS thickness and guide field strengths.


This paper is organized as follows: the simulation code and setup are described in \S\ref{sec:setup}; the result of a simulation with an ion-scale CS that reveals X-line spreading slower than the guide-field Alfv\'en speed $V_{Ag}$, qualitatively consistent with recent magnetopause observations \citep{Zou:2018}, is presented in \S\ref{sec:bg2}; the result of a simulation with a sub-ion-scale CS showing Alfv\'enic X-line spreading at $V_{Ag}$ is presented in \S\ref{sec:bg4}, including analysis on the thinning and extending of the CS that leads to subsequent reconnection onset, and an Alfv\'enic signal that causes the CS thinning; the role of the collisionless tearing instability, the tearing growth rate and measurements of four simulations are discussed in \S\ref{sec:gamma_sim}; in \S\ref{sec:diss}, the present work is compared with previous numerical studies; implications of this work for the magnetopause and turbulent magnetosheath are discussed, and the main findings are summarized in \S\ref{sec:imp}.

\section{Simulation Setup}\label{sec:setup}
Large-scale simulations are performed with the electromagnetic particle-in-cell code {\it VPIC} \citep{bowers09a}. The initial asymmetric current sheet \citep{yhliu18b,yhliu15b,hesse13a,aunai13b,pritchett08a} is given by the magnetic profile ${\bf B}_0=B_0[(0.5+S)\hat{\bf x} + b_g\hat{\bf y}]$ with $S=\mbox{tanh}[(z-z_0)/L]$, where the guide field strength is $B_g/B_0$ = $b_g$ and $z_0$ is the shift of the current sheet from $z=0$. This profile gives asymptotic magnetic fields $B_{2x0} = 1.5B_0$ and $B_{1x0} = -0.5B_0$ where the subscripts ``1'' and ``2'' correspond to the magnetosheath and magnetosphere sides, respectively. The shift $z_0$ allows room for the reconnected asymmetric field to bulge into the weaker-field side. The plasma has a density profile $n=n_0[1-(S+S^2)/3]$ giving upstream densities of $n_2=n_0/3$ and $n_1=n_0$. The uniform total temperature $T=3B_0^2/(8\pi n_0)$ consists of contributions from ions and electrons with a ratio of $T_i/T_e=5$ to simulate typical magnetopause conditions. The mass ratio is $m_i/m_e=25$. The ratio of the electron plasma to gyro- frequencies based on $B_0$ is $\omega_{pe}/\Omega_{ce}=4$ where $\omega_{pe}\equiv(4\pi n_0 e^2/m_e)^{1/2}$, $\Omega_{ce}\equiv eB_0/m_e c$. Velocities, spatial scales, time, densities and magnetic fields, unless otherwise stated, are normalized to the Alfv\'en speed (based on $B_0$) $V_{A0}\equiv B_0/(4\pi n_0 m_i)^{1/2}$, the ion inertia length $d_i\equiv c/\omega_{pi}$, the ion gyro-frequency $\Omega_{ci}$, $n_0$ and $B_0$, respectively.

Four simulations are performed. Two have a thicker, ion-scale initial current sheet (CS) with a half-thickness $L= 0.8 d_i$ and guide field strengths $b_g$ = 2 and 1, labeled as {\it Ibg2} and {\it Ibg1}, respectively. The other two simulations have a thinner, sub-ion-scale CS with a half-thickness $L= 0.4 d_i$ and $b_g$ = 4 and 2, labeled as {\it Sbg4} and {\it Sbg2}, respectively. The domain size of {\it Ibg2} and {\it Ibg1} is $L_x \times L_y \times L_z=128d_i \times 256d_i \times 24d_i$ and $256d_i \times 256d_i \times 24d_i$, respectively. Both has a shift $z_0=3 d_i$. The number of grid cells is $(n_x,n_y,n_z)=(4096,4096,1024)$ and $(4096,4096,384)$, respectively. For the thinner CS simulations, both has a domain size of $128d_i \times 256d_i \times 12d_i$ and a shift of $z_0= d_i$. The number of grid cells $(n_x,n_y,n_z)$ is $(8192,4096,1024)$ and $(4096,4096,512)$ for {\it Sbg4} and {\it Sbg2}, respectively. There is a total of approximately 0.9, 1.3, 0.9 and 0.4 trillion particles in the {\it Ibg2}, {\it Ibg1}, {\it Sbg4} and {\it Sbg2} simulation, respectively. For all simulations, the boundary conditions are periodic both in the $x$- and $y$-directions; in the $z$-direction, they are conducting for fields and reflecting for particles.  We adopt the same methodology as in \cite{yhliu15b}, using localized perturbations in both $x$- and $y$-directions to initiate reconnection near the center of the domain at $(x,y,z)=(0,0,z_0)$.


\begin{figure}[htb]
\centering

\hbox{ \hfill \resizebox{3.5in}{!}{\includegraphics[scale=1.,trim=6cm 8.5cm 4cm 6cm, clip=true]{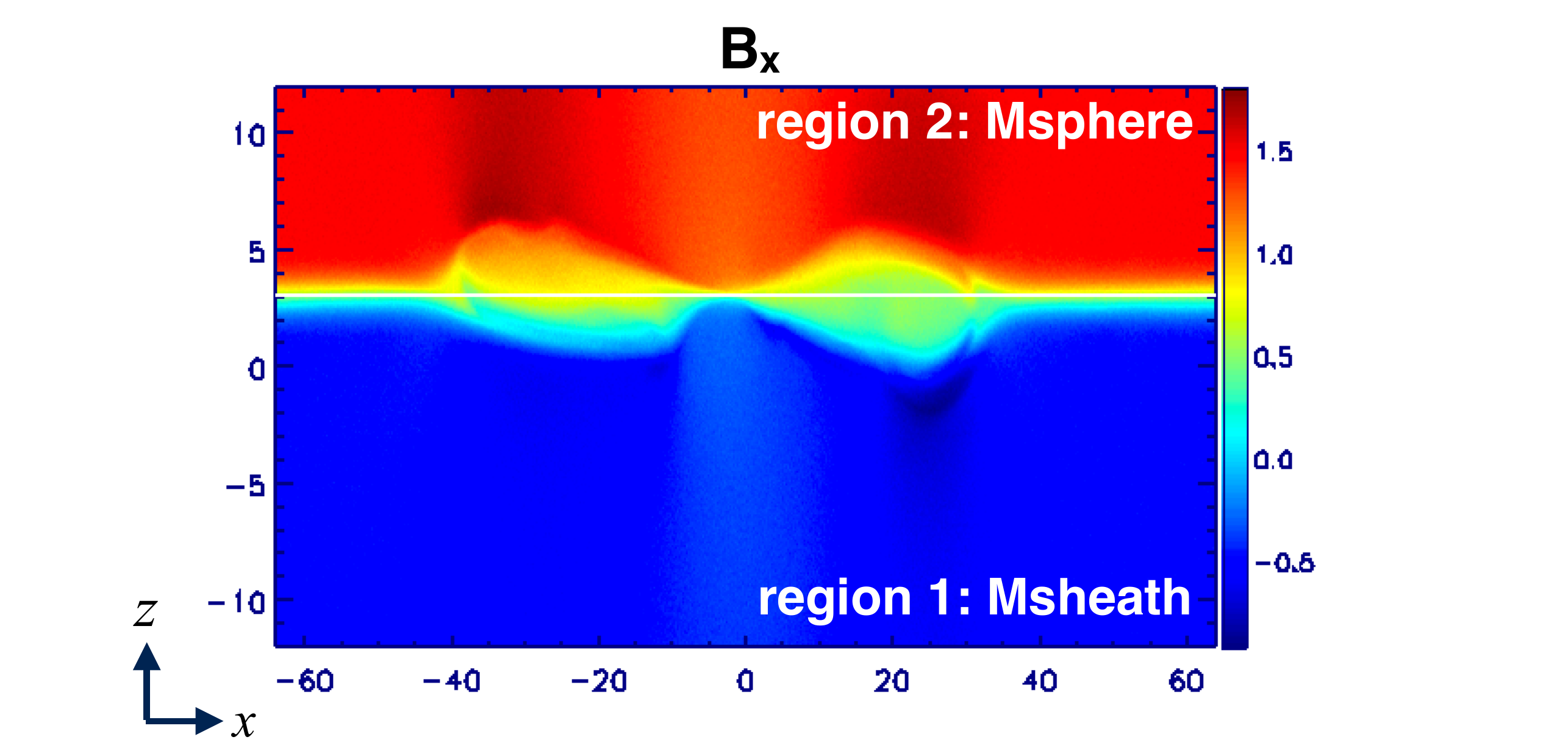}} \hfill  }
\caption{ \label{fig:bg2-bx} Run \emph{Ibg2}: A cut of $B_x$ at $y$=0 at late time $\Omega_{ci}t$ = 128, showing a fully developed reconnection state. The horizontal white line cuts through the X-line. }
\end{figure}


\begin{figure*}[htb]
\centering

\hbox{ \hfill \resizebox{6.in}{!}{\includegraphics[scale=1.,trim=0.1cm 8.5cm 0.1cm 5cm, clip=true]{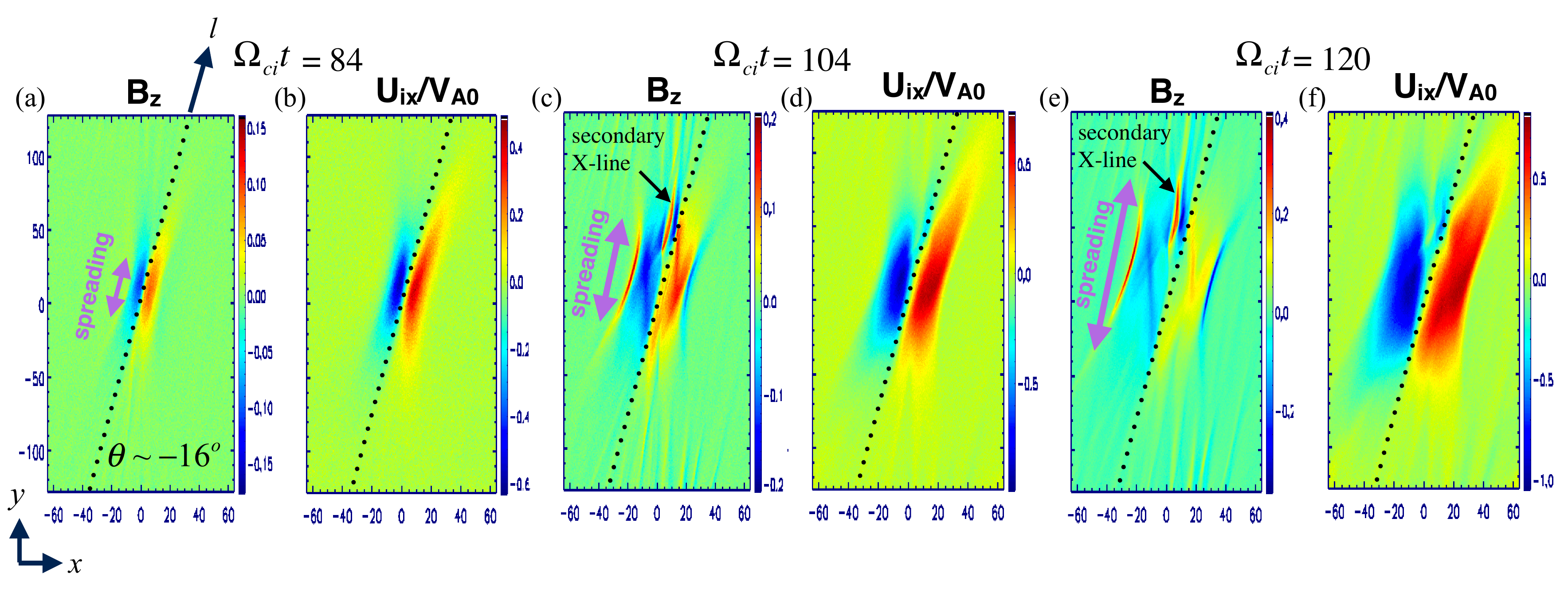}} \hfill  }
\caption{ \label{fig:bg2-bz-uix-sp} Run \emph{Ibg2}: X-line spreading observed in the (normalized) reconnected magnetic field $B_z/B_0$ and ion outflow $U_{ix}/V_{A0}$ at $\Omega_{ci}t$ = 84, 104, 120. The X-line orientation, labelled as the $l$ axis and indicated by dotted black lines, makes an angle of $\theta$ $\sim$ -16$^\circ$ with the $y$ axis. Different scales are used for the colorbars to best illustrate X-line spreading. }
\end{figure*}


\section{Simulation with X-line Spreading speed below $V_{Ag}$}\label{sec:bg2}
We first present the result from one of the simulations in which X-line spreading is sub-Alfv\'enic, {\it i.e.} below $V_{Ag}$. This result is qualitatively consistent with recent magnetopause observations \citep{Zou:2018}. Fig.~\ref{fig:bg2-bx} shows a cut of $B_x$ on the $x$-$z$ plane at $y$=0 at late time when the X-line is well formed and reconnection fully developed. The upper part with asymptotic $B_x$>0 represents the magnetosphere side and the lower part with asymptotic $B_x$<0 represents the magnetosheath side of the magnetopause. To examine the system on the $x$-$y$ plane, we take a cut at $z$ = $z_c$ (indicated by the solid white line) such that $B_x(z_c)$ = 0.6, approximately halfway between $B_{1x0}$ and $B_{2x0}$, representing the close vicinity of the diffusion region. We note that $z_c$ stays approximately constant in time. We can neglect the movement of the X-line along $z$ by observing that the (intense) CS stays confined to $\sim$1 $d_i$ along $z$ through late time, a negligible displacement compared to the much larger displacement of the X-line on the $x$-$y$ plane. The extending, reconnecting CS is then effectively sampled on the $x$-$y$ plane at $z_c$ for measuring the X-line spreading.

Fig.~\ref{fig:bg2-bz-uix-sp} shows the reconnected magnetic field $B_z$ and ion outflow $U_{ix}$ on the sampled $x$-$y$ plane at different times. Because of asymmetric upstream conditions, the X-line orientation makes an angle $\theta\sim$ -16$^\circ$ with the $y$ axis \citep{yhliu18b,yhliu15b}. This orientation is referred to as the $l$ axis. At $\Omega_{ci}t$ = 84, shortly after reconnection starts, both $B_z$ and $U_{ix}$ are localized, having a short extent in the $l$ direction. The reversal of $B_z$ and $U_{ix}$ indicates the position of the X-line. The localized X-line extends over time, as seen at $\Omega_{ci}t$ = 104 and 120. At $\Omega_{ci}t$ = 104, a secondary X-line near $y\sim$ 50 $d_i$ (indicated by a black arrow) emerges from secondary tearing instability, as shown in $B_z$. This secondary X-line also extends in time, as observed at $\Omega_{ci}t$ = 120.


\begin{figure*}[htb]
\centering

\hbox{ \hfill \resizebox{5.5in}{!}{\includegraphics[scale=1.,trim=0.5cm 5.5cm 3cm 4cm, clip=true]{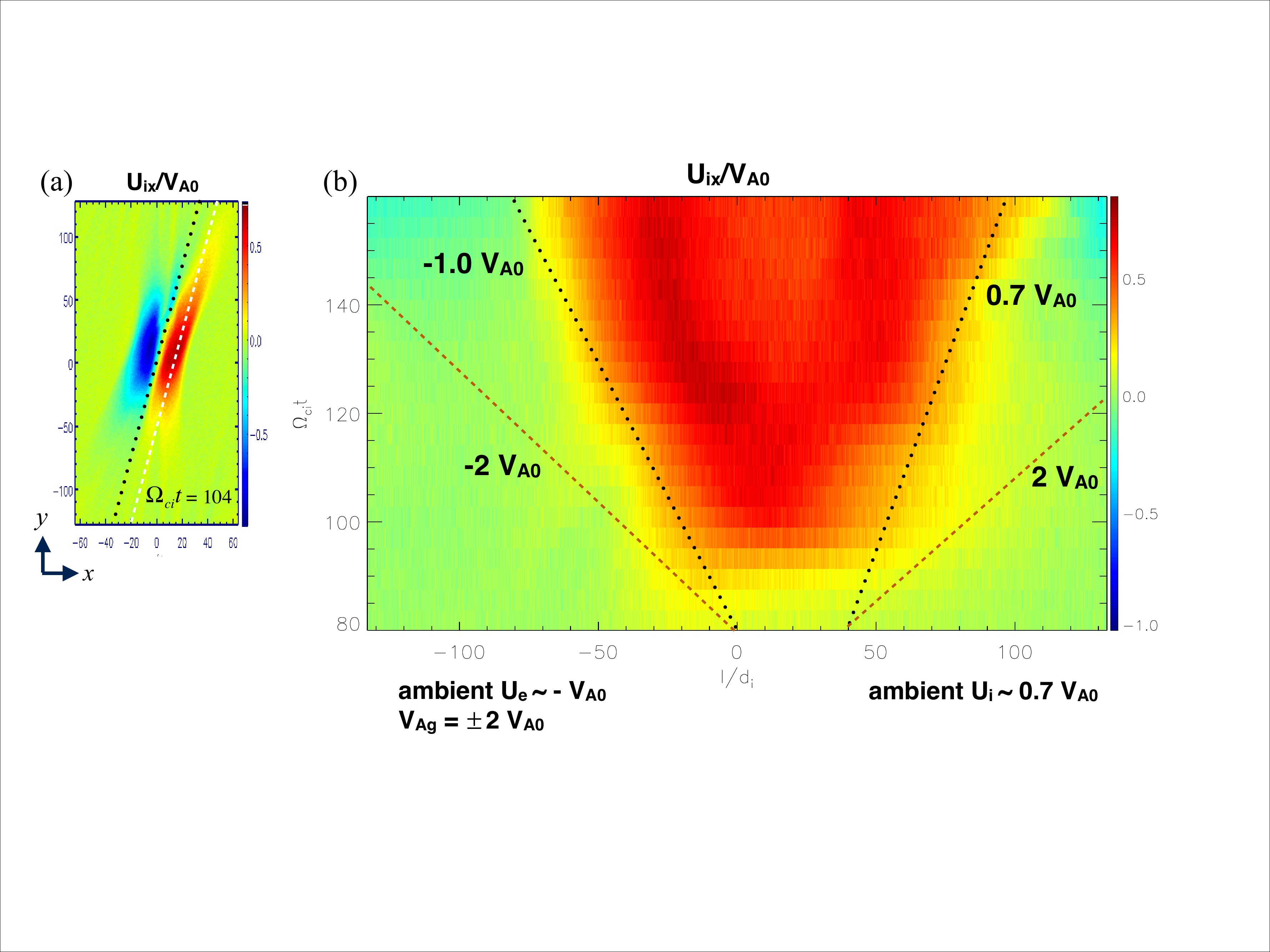}} \hfill 
}
\caption{ \label{fig:bg2-sp} Run {\it Ibg2}: X-line spreading measured from the ion outflow $U_{ix}$: (a) a cut of $U_{ix}$ across the initial X-line on the $x$-$y$ plane at $\Omega_{ci}t$=104. The orientation of the X-line is indicated by a dotted black line (through $(x,y)$=(0,0)) and spreading of $U_{ix}$ is sampled at the dashed white line for a timestack plot. (b) A timestack plot of sampled $U_{ix}$ across the right dashed white line.    }
\end{figure*}


Plotted in Fig.~\ref{fig:bg2-sp} is the measurement of X-line spreading from the spreading of the ion outflow. In (a), the $U_{ix}$>0 region (red) is sampled across the dashed white line parallel to the $\hat{l}$ axis to measure the extension of the ion outflow region. Timestacks of the sampled $U_{ix}$ is plotted in (b).

The $l$ extend of $U_{ix}$ is short at earlier time, e.g., $\Omega_{ci}t$ = 100, and has extended significantly at late time. The slope of the Alfv\'enic outflow region with $U_{ix}\sim$ 0.5 $V_{A0}$ is the spreading speed of the extending outflow region, a proxy for the X-line spreading speed. Along the $+\hat{l}$ direction, the spreading speed is measured to be $V_{s+}\sim$ 0.7 $V_{A0}$, which is comparable to the ambient ion drift speed $U_i\sim$ 0.7 $V_{A0}$. Along the $-\hat{l}$ direction, the spreading speed is measured to be $V_{s-}\sim$ -1.0 $V_{A0}$, comparable to the ambient electron drift speed $U_e\sim$ -$V_{A0}$; \S\ref{bg2-ue} illustrates the determination of $U_e$ and $U_i$. Hence, the X-line spreading is consistent with the ion/electron drift speeds \citep{TKMNakamura12}. In contrast, both $V_{s\pm}$ is about twice slower than the guide-field Alfv\'en speed $V_{Ag}$ = $\pm$2 $V_{A0}$. This reveals that even though $V_{Ag}$ is significantly higher than the ambient ion/electron drift speeds, it does not mediate the X-line spreading. This simulation is qualitatively consistent with the recent magnetopause event \citep{Zou:2018} in which the X-line spreads much more slowly than the expected $V_{Ag}$ \citep{Shepherd:12}.


\begin{figure}[htb]
\centering

\hbox{ \hfill \resizebox{4.5in}{!}{\includegraphics[scale=1.,trim=2cm 0.1cm 2cm 0.1cm, clip=true]{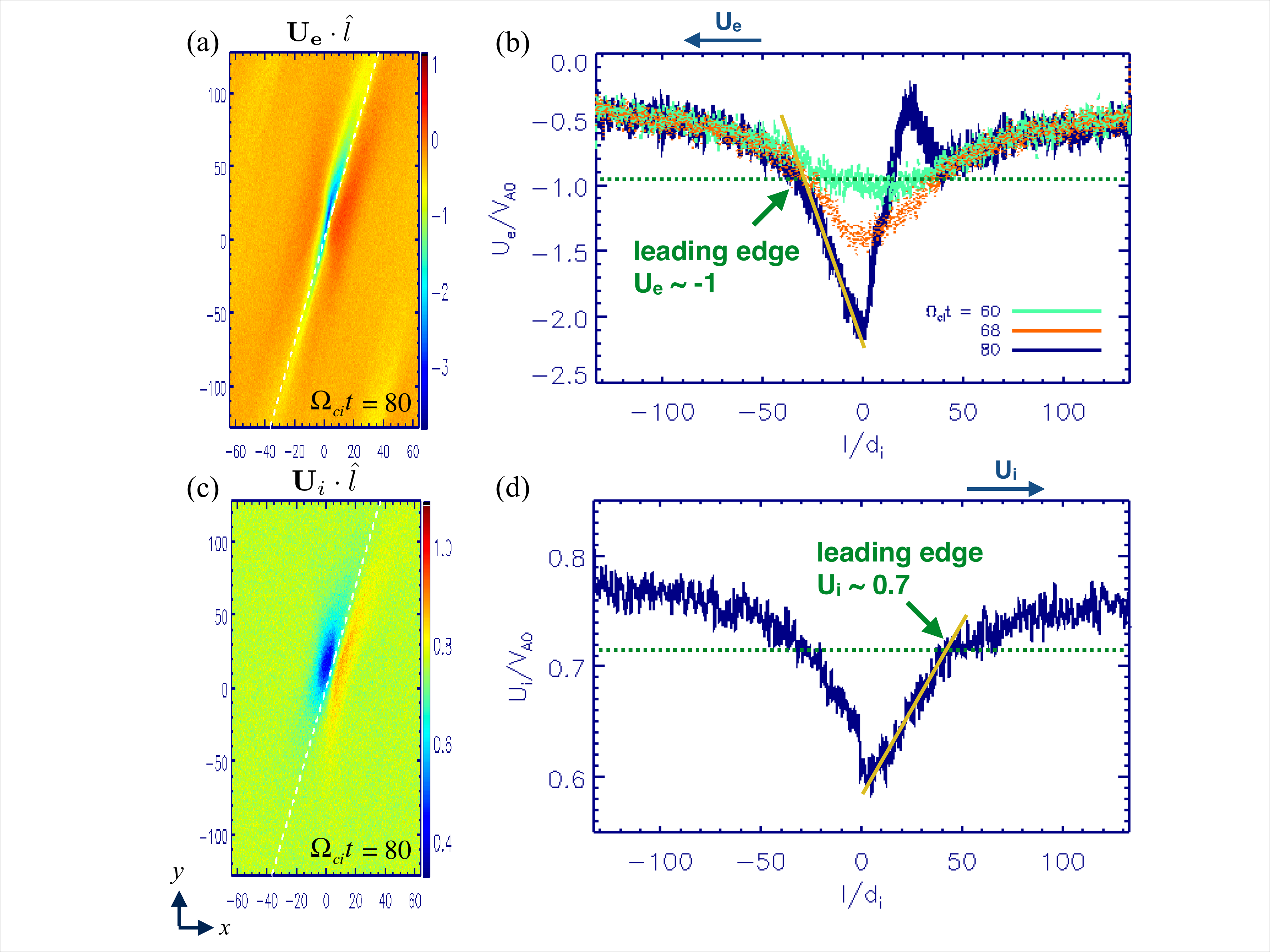}} \hfill 
}
\caption{ \label{fig:bg2-ue} Run {\it Ibg2}: (a) The electron drift speed directed along the $\hat{l}$ axis at $\Omega_{ci}t$ = 80. Same format as Fig.~\ref{fig:bg2-bz-uix-sp}. Cuts along the dashed white line at various times are plotted in (b), showing the development of high electron drift speed. The ambient or \emph{leading edge} $U_e$ is depicted. (c) The ion drift speed directed along the $\hat{l}$ axis and (d) a cut along the dashed white line at $\Omega_{ci}t$ = 80, with the ambient or \emph{leading edge} $U_i$ depicted.    }
\end{figure}

\subsection{Ambient ion/electron drift speeds}\label{bg2-ue}
The ambient electron and ion drift speeds, $U_e$ and $U_i$, are determined as follows. Fig.~\ref{fig:bg2-ue} (a) shows a cut of the $\hat{l}$-directed electron drift speed, $\mathbf{U_e}\cdot\hat{\mathbf{l}}$, on the $x$-$y$ plane at $\Omega_{ci}t$ = 80. Plotted in (b) is time evolution of $\mathbf{U_e}\cdot\hat{\bf l}$ cuts along the dashed white line in (a), which is aligned with $\hat{l}$. For simplicity, we define $U_e\equiv\mathbf{U_e}\cdot\hat{\bf l}$. The background $U_e$ is $\sim$ -0.5$V_{A0}$. An increasing electron drift develops within a localized region from $\Omega_{ci}t$ = 60 (green), before reconnection, to $\Omega_{ci}t$ = 80 (blue), the start of reconnection. Characterized by a steep gradient (depicted by a yellow line in (b)), the localized high electron drift region indicates the reconnection region. Immediately outside the left edge (-$\hat{l}$ side) of the localized reconnection region is defined as the {\it leading edge} where the ambient $U_e$ is measured. It is followed by a gradual transition, which remains nearly stationary in time (as seen in the overlapping $U_e$ cuts), and eventually drops to the background value of $\sim$ -0.5$V_{A0}$. $U_e$ at the leading edge carries reconnection signals of the localized reconnection region to the immediate ambient plasma (see \S\ref{sec:diss} for a picture of the X-line spreading mediated by the current carriers proposed by \cite{TKMNakamura12}). Note that the leading edge $U_e$ stays at $\sim -V_{A0}$ on average throughout the X-line spreading. We note that in comparison, the peak $U_e$ ($\sim -2V_{A0}$) at the center of the reconnection region apparently cannot explain the measured X-line spreading speed.


The same procedure is used to determine the ambient ion drift speed. Fig.~\ref{fig:bg2-ue}(c) shows the $\hat{l}$-directed ion drift speed, $\mathbf{U_i}\cdot\hat{\mathbf{l}}$, on the $x$-$y$ plane and (d) a cut along the dashed white line, at $\Omega_{ci}t$ = 80. For simplicity, we define $U_i\equiv\mathbf{U_i}\cdot\hat{\bf l}$. Note that $U_i$ is not significantly modified in the reconnection region that is indicated by a steep-gradient drop from its background value. Immediately outside the right edge (+$\hat{l}$ side) of the reconnection region is the leading edge where the ambient $U_i$ is measured. The measured ambient $U_i$ stays at $\sim$ 0.7$V_{A0}$ on average throughout the X-line spreading, and is close to the background value of $\sim$ 0.8$V_{A0}$. 

\section{Simulation with X-line Spreading speed at $V_{Ag}$}\label{sec:bg4}

\begin{figure*}[htb]
\centering

\hbox{ \hfill \resizebox{6.in}{!}{\includegraphics[scale=1.,trim=1cm 8cm 1cm 3cm, clip=true]{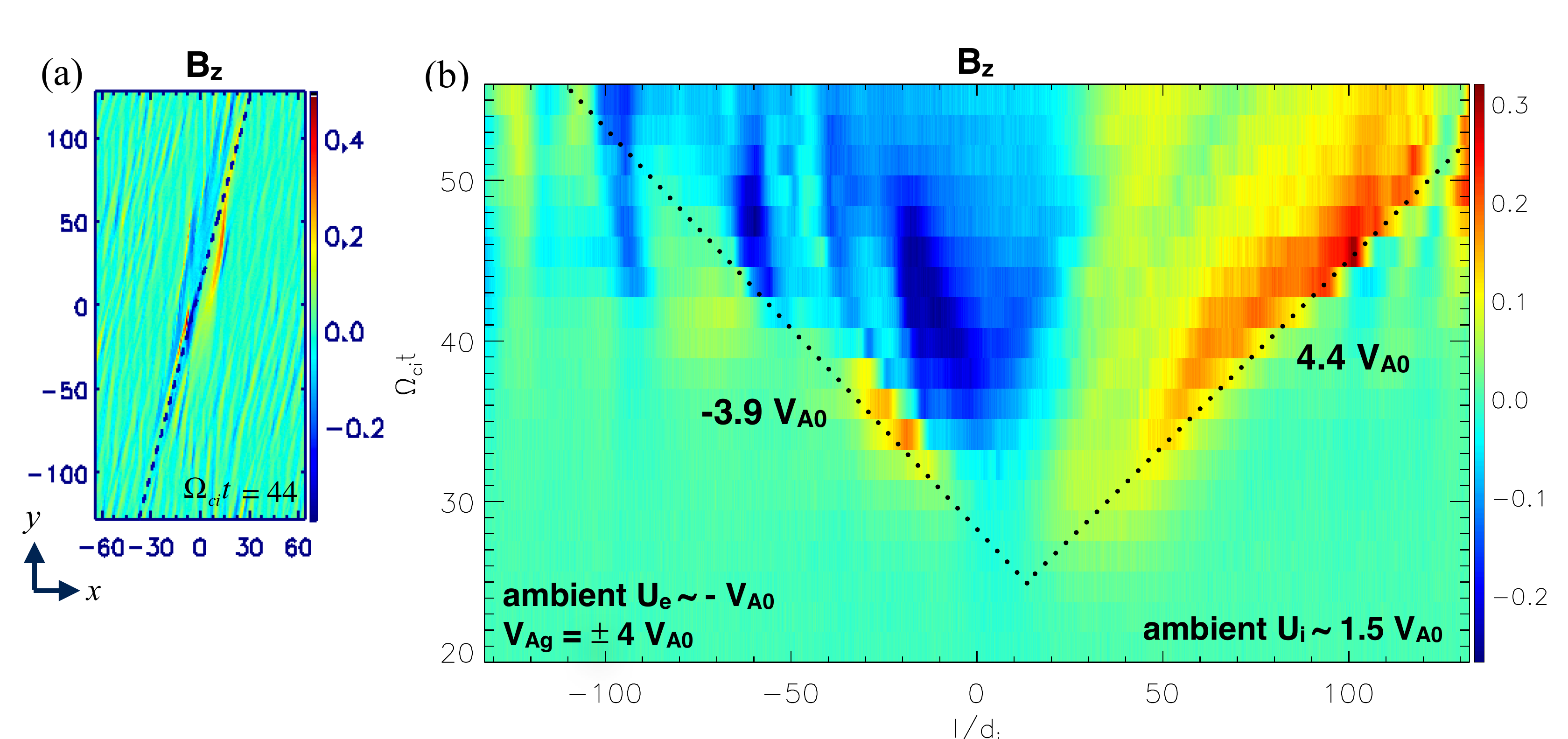}} \hfill 
}
\caption{ \label{fig:bg4-sp} Run {\it Sbg4}: X-line spreading measured from the reconnected magnetic field $B_z$: (a) a cut of $B_{z}$ on the sampled $x$-$y$ plane at $\Omega_{ci}t$=44 (during reconnection), showing $B_z$ reversal. The orientation of the X-line is denoted by the dashed black line. (b) A timestack plot of $B_z$ sampled across the dashed black line for measuring the speed of $B_z$ spreading. }
\end{figure*}

We now present results from the {\it Sbg4} simulation. Plotted in Fig.~\ref{fig:bg4-sp} (a) is a cut of $B_z$ on the sampled $x$-$y$ plane at $\Omega_{ci}t$ = 44, a time during reconnection. The same procedure of sampling as in \S\ref{sec:bg2} is used.

Despite more structured due to the emergence of secondary X-lines, the reversal of $B_z$ of the primary X-line can be observed over a broad region of which $B_z$<0 (light blue) is approximately on the left and $B_z$>0 (yellow) is on the right of the X-line orientation (dashed line). The X-line orientation is measured to be $\theta\sim$-15$^\circ$ (dashed black line), which aligns well with the reversal of $B_z$ for the $y$<0 side; for the $y$>0 side $\theta\sim$-12$^\circ$ is possible while both $\theta$'s measure comparable spreading speeds to within $\sim$20\%. $B_z$ reversals of narrower, secondary X-lines are also present. The dashed black line therefore samples $B_z$ from both the primary and secondary X-lines. Plotted in (b) is a timestack plot of $B_z$ sampled. 



The spreading speeds are estimated to be $V_{s+}$$\sim 4.4 V_{A0}$ and $V_{s-}$$\sim -3.9 V_{A0}$, in close agreement with the guide-field Alfv\'en speed $V_{Ag}$ = $\pm$4 $V_{A0}$. A timestack plot using the ion outflow $U_{ix}$ (not shown) has qualitatively the same features as $B_z$, showing  $U_{ix}$ spreading in the $\pm\hat{l}$. It measures $V_{s\pm}$ comparable to those obtained from the $B_z$ timestack plot. We note that, in contrast to run {\it Ibg2}, the measured spreading speeds, $V_{s-}$ and $V_{s+}$, are much higher than the ambient electron drift speed $U_e\sim$ -$V_{A0}$ and the ambient ion drift speed $U_i\sim$ 1.5$V_{A0}$, respectively, showing that the ion/electron drifts does not mediate Alfv\'enic X-line spreading at $V_{Ag}$ observed here. The question is then what causes the Alfv\'enic X-line spreading in a thinner CS.

\subsection{CS Thinning and X-line Formation}\label{bg4-CS}

\begin{figure}[htb]
\centering

\hbox{ \hfill \resizebox{4.in}{!}{\includegraphics[scale=3.,trim=3cm 7.5cm 3cm 3cm, clip=true]{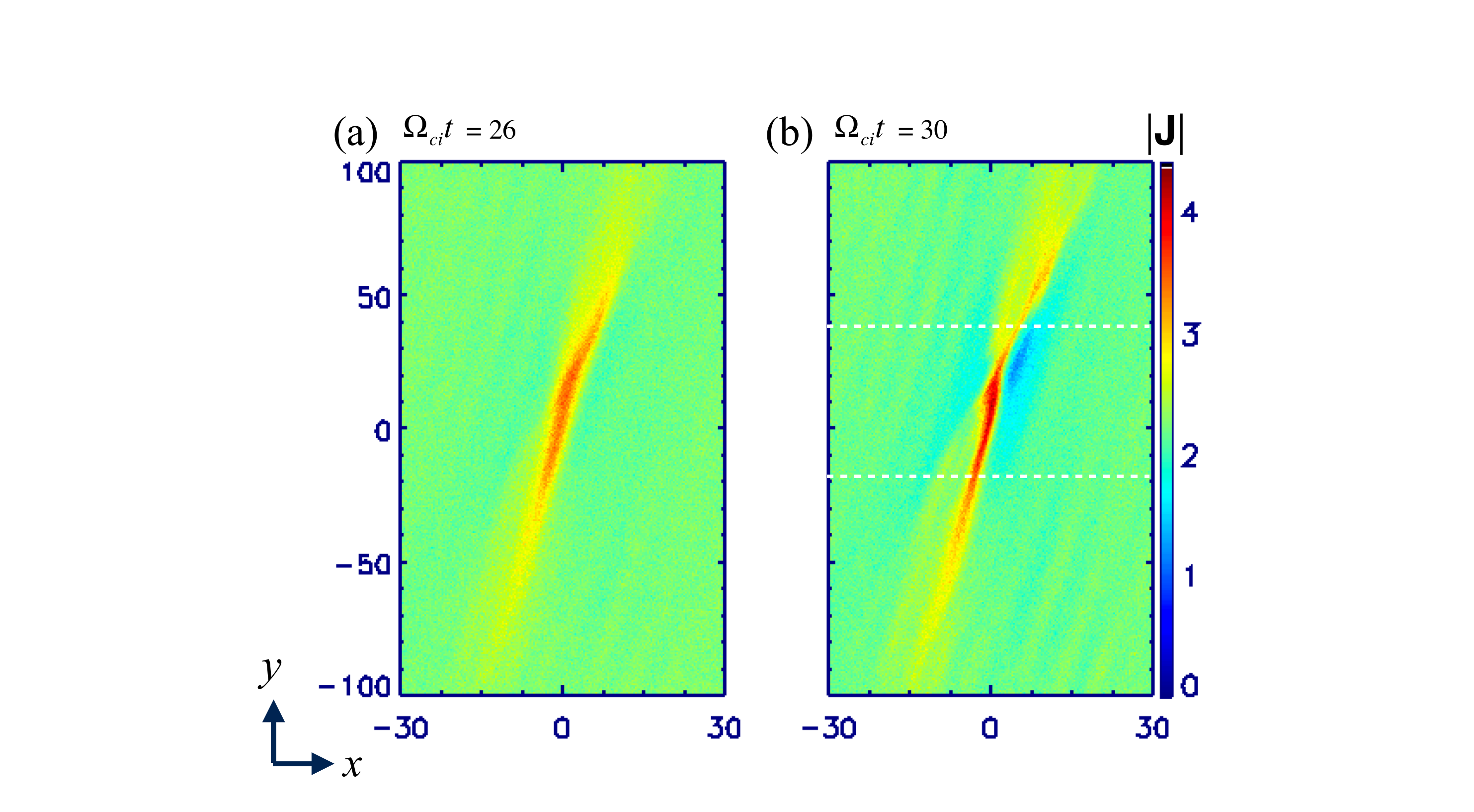}} \hfill 
}
\caption{ \label{fig:bg4-absJ-xy} Run \emph{Sbg4}: Total current density $|\mathbf{J}|$ on the $x$-$y$ plane at $\Omega_{ci}t$= (a) 26 and (b) 30.}
\end{figure}

To understand the underlying physics of X-line spreading at $V_{Ag}$, we first examine the development of the thinner CS. Fig.~\ref{fig:bg4-absJ-xy} shows a cut of the total current density $|\mathbf{J}|$ on the $x$-$y$ plane (similar to Fig.~\ref{fig:bg4-sp}(a)) at $\Omega_{ci}t$= (a) 26 and (b) 30, which are approximately the times of reconnection onset when finite $B_z$ of amplitude $\sim$0.1$B_0$, typical of fast reconnection, emerges (see Fig.~\ref{fig:bg4-sp}(b)). \emph{Onset} here refers to the onset of fast reconnection. Here we zoom into the central region of the $x$-$y$ plane to illustrate more clearly the evolution of $|\mathbf{J}|$.

Several key features are observed. First, the CS thins over time. Second, the thinning of the CS extends in both the $\pm\hat{l}$ directions (that are consistent with the fastest tearing growing modes). Third, this CS thinning proceeds fast, on a short time scale of a few $\Omega_{ci}^{-1}$


\begin{figure}[htb]
\centering

\hbox{ \hfill \resizebox{4.in}{!}{\includegraphics[scale=3.,trim=3cm 6cm 3cm 3cm, clip=true]{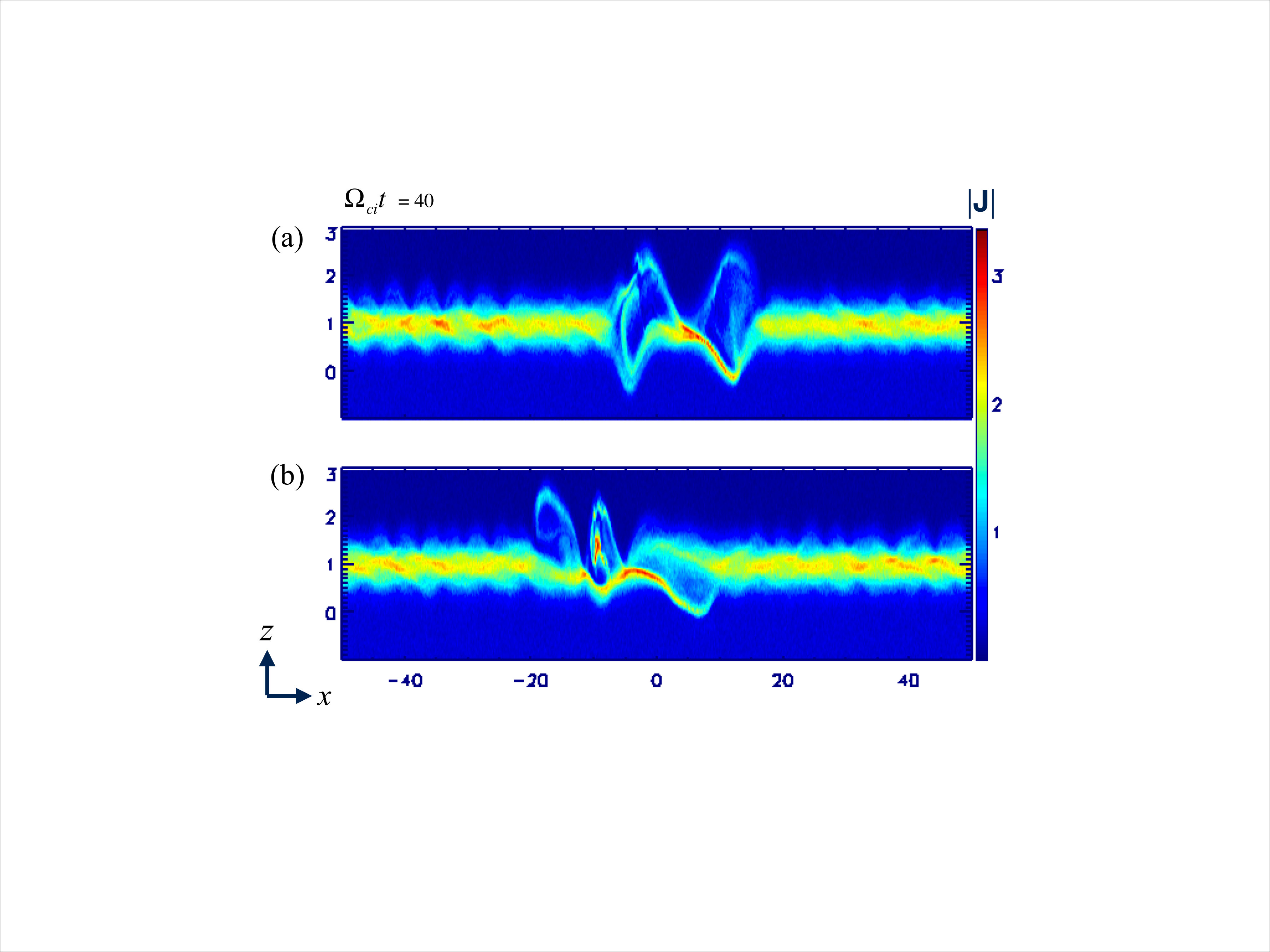}} \hfill 
}
\caption{ \label{fig:bg4-absJ-xz} Run \emph{Sbg4}: Cut of the total current density $|\mathbf{J}|$ on the $x$-$z$ plane along the (a) top and (b) bottom dashed lines in Fig.~\ref{fig:bg4-absJ-xy}(b), at a later time of $\Omega_{ci}t$=40, showing the formation of an X-line.}
\end{figure}


Progressive thinning of the CS is expected to onset reconnection, leading to the formation of an X-line. To examine if reconnection onsets following the thinning of the CS that extends towards the $\pm\hat{l}$ directions, we plot in Fig.~\ref{fig:bg4-absJ-xz} a cut of $|\mathbf{J}|$ on the $x$-$z$ plane at locations denoted by the two dashed lines in Fig.~\ref{fig:bg4-absJ-xy}(b) at a later time $\Omega_{ci}t$=40. At both locations, an X-line is forming near $x=0$, showing that reconnection has onset, as expected. Thus, the extending of CS thinning leads to reconnection onset further away from the center of the domain. X-line spreading can then be understood as the continuous onset of reconnection along the $\pm\hat{l}$ directions.


\subsection{Signal Development during the Pre-onset phase}\label{bg4-signal}
\begin{figure}[htb]
\centering

\hbox{ \hfill \resizebox{6.in}{!}{\includegraphics[scale=.01,trim=0.1cm 0.5cm 0.1cm 0.5cm, clip=true]{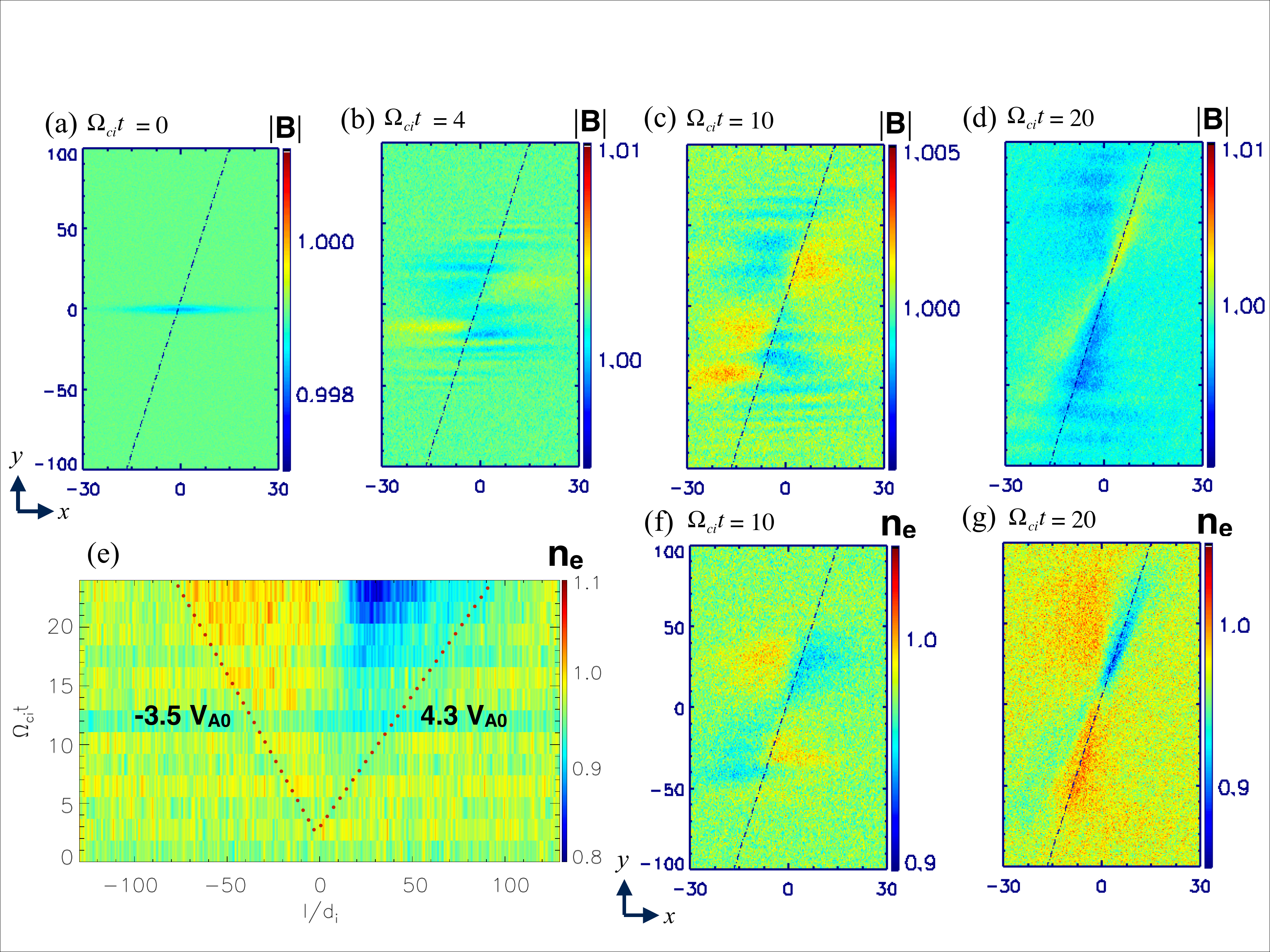}} \hfill 
}
\caption{ \label{fig:bg4-signal} Run \emph{Sbg4}: Time evolution of (a)-(d) the total magnetic field strength $|\mathbf{B}|$ normalized to its initial value $|\mathbf{B}(t$=$0)|$ and (f)-(g) electron density $n_e$ on the $x$-$y$ plane using the same format as Fig.~\ref{fig:bg4-absJ-xy}, showing the development and propagation of a signal arising from the initial perturbations in the magnetic field. (e) Timestack plot of $n_e$ sampled along the dashed line during the pre-onset phase similar to Fig.~\ref{fig:bg4-sp}(b).  }
\end{figure}

To understand what causes CS thinning and extending, which lead to reconnection onset, we investigate the evolution of the system prior to reconnection onset, {\it i.e.}, the pre-onset phase at $\Omega_{ci}t<25$. Here we focus on the total magnetic field and electron density. Fig.~\ref{fig:bg4-signal} shows (a)-(d) the time evolution the total magnetic field $|\mathbf{B}|$ normalized to the initial value and the corresponding (f)-(g) electron density $n_e$ on the $x$-$y$ plane prior to onset, using the same format as Fig.~\ref{fig:bg4-absJ-xy}. Plotted in (e) is the timestack plot of $n_e$ sampled along the dashed line during the pre-onset phase using the same procedure as Fig.~\ref{fig:bg4-sp}(b).


An Alfv\'enic signal consistent with kinetic Alfv\'en waves develops from initial perturbations in the magnetic field, and propagates at approximately the local Alfv\'en speed $\sim V_{Ag}$ bidirectionally. Fig.~\ref{fig:bg4-signal} illustrates details of its development.
At (a) $\Omega_{ci}t$=0, localized initial perturbations in the $x$ and $z$ components of the magnetic field are imposed. They amplify over time as seen at (b) $\Omega_{ci}t$=4. Amid the shorter-scale initial perturbations emerges a signal that is characterized by a long length scale parallel to the orientation of the later-formed X-line (represented by the dashed lines). The parallel length scale of the signal becomes macroscopic, comparable to $L_y$, by (d) $\Omega_{ci}t$=20. In addition, short perpendicular scales in the $x$ and $z$ (not shown) directions also develop. The perpendicular scales are estimated to be $k_x\rho_i\sim$0.1, and $k_z\rho_i\sim$1 given the sub-$d_i$ thickness of the CS. 
The signal is compressive, as characterized by its amplitude in the electron density $n_e$ (despite no initial density perturbations imposed) in (f) and (g) corresponding to the long-parallel-scale magnetic field $|\mathbf{B}|$ perturbations. Note that $|\mathbf{B}|$ and $n_e$ perturbations of the signal are anti-correlated, presumably to maintain a state of pressure balance. A Timestack plot of $n_e$ (e) indicates that the signal propagates bidirectionally along the later-formed X-line orientation at approximately the local Alfv\'en speed, which is dominated by the guide field near the center of the CS and hence is $\sim V_{Ag}=\pm$ 4 $V_{A0}$. The characteristics of this signal are consistent with kinetic Alfv\'en waves.

The Alfv\'enic signal perturbs and thins the CS as it propagates, extending CS thinning across the system, which results in the CS thinning and extending as seen in Fig.~\ref{fig:bg4-absJ-xy}. A timestack plot of $|\mathbf{J}|$ (not shown) reveals that the speed of the extending thinning CS during this pre-onset phase is $\sim\pm$4.6 $V_{A0}$, consistent with the signal speed. The signal and thinning CS also share the same orientation, as expected.

Note that in run {\it Ibg2}, an Alfv\'enic signal also develops, but the resulted CS thinning is much weaker and slower than that in this run because of the thicker equilibrium CS there. Hence, the signal does not mediate X-line spreading in that case. This indicates that the time scale of CS thinning and the resulted reconnection onset, expected to be strongly dependent on the CS thickness, is important to whether Alfv\'enic X-line spreading can take place.




\begin{figure}[htb]
\centering

\hbox{ \hfill \resizebox{3.in}{!}{\includegraphics[scale=1.,trim=9cm 4cm 11cm 4cm, clip=true]{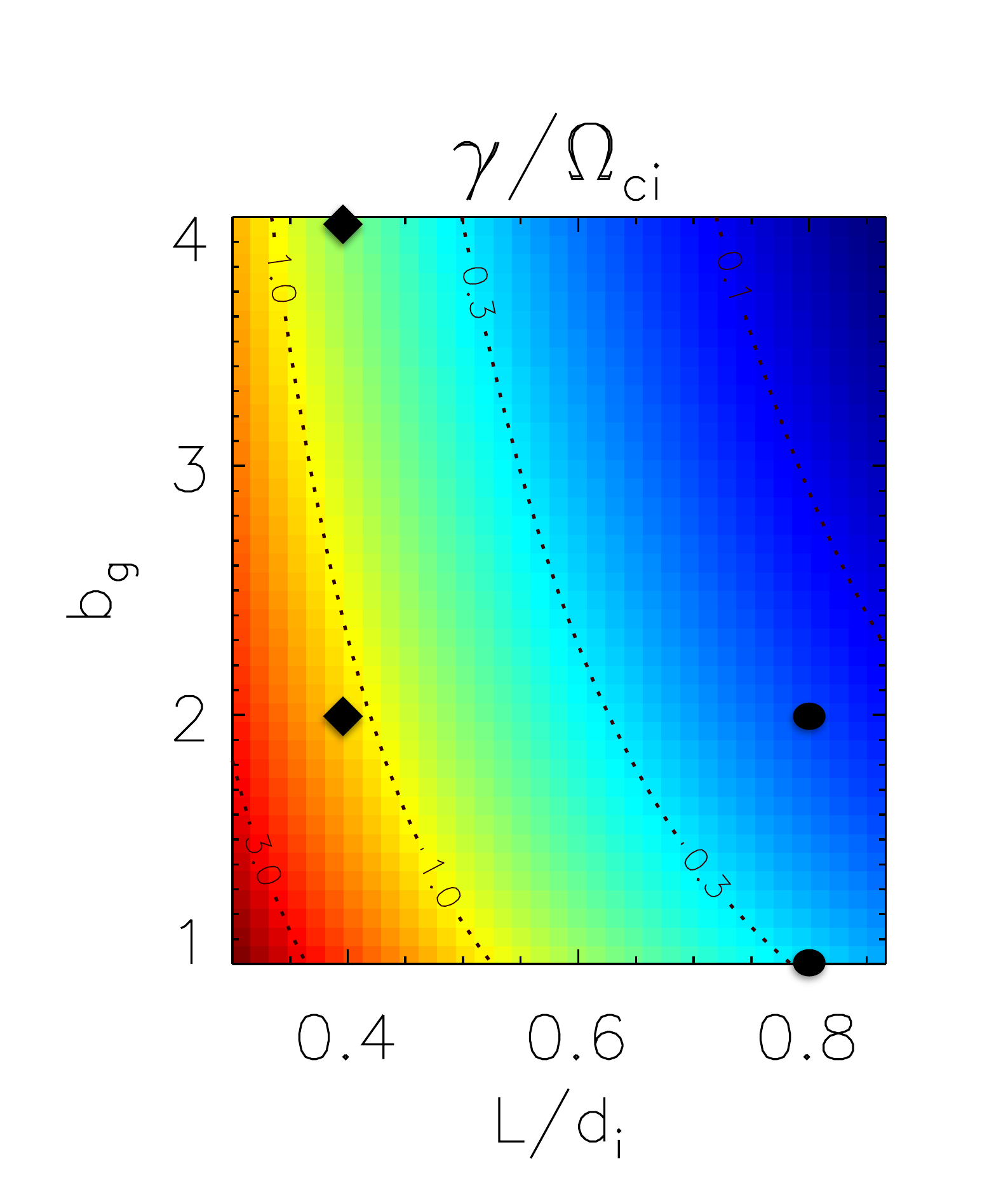}} \hfill 
}
\caption{ \label{fig:gamma} Growth rate of the dominant collisionless tearing mode normalized to $\Omega_{ci}$ as a function of CS thickness and guide-field strength $\gamma (L/d_i,b_g)/\Omega_{ci}$ in logarithmic scale. Dots denote simulations in lower $\gamma$ region with $\gamma/\Omega_{ci}$ of order 0.1. Diamonds denote simulations with $\gamma/\Omega_{ci}\sim$1. }
\end{figure}

\section{Collisionless tearing growth rate}\label{sec:gamma_sim}

We have shown that X-line spreading is based on the spreading of reconnection onset. The spreading speed has a clear dependence on the CS thickness. The collisionless tearing instability \citep{daughton11a} that is known to initiate reconnection, and found to be the dominant instability in 3D guide-field reconnection \citep{yhliu18b,yhliu13a} naturally explains the CS thickness dependence of the different X-line spreading behaviours, Alfv\'enic or sub-Alfv\'enic, in our simulations; its growth rate is strongly dependent on the CS thickness. Note that the orientation of the thinning CS and resulted X-line in our simulations is consistent with the fastest growing modes of this instability, as expected. To estimate the time scale of reconnection onset, we calculate the linear growth rate of collisionless tearing \footnote{Note that to fully describe reconnection onset, nonlinear theory of the collisionless tearing instability is required. Although such a theory is not yet available, it is expected to strongly depend on the CS thickness as well.}

The linear collisionless tearing growth rate as a function of guide field strength $b_g$ and CS (half) thickness $L$, based on simulation parameters and magnetic profile, is plotted in Fig.~\ref{fig:gamma} in logarithmic scale. The analytical form of $\gamma$ was previously derived \citep{yhliu18b}, and the derivation is given in \S\ref{appx} Appendix. Here $\gamma$ is taken as the maximum growth rate over all oblique angles $\theta$ for an arbitrary wavevector ${\bf k}=k_x\hat{{\bf x}}+k_y\hat{{\bf y}}$ with $\tan\theta \equiv k_y/k_x$ of an oblique tearing mode.

The functional dependence of $\gamma$ on $L$ and $b_g$ can be approximated as: 
\begin{equation}
\gamma/\Omega_{ci} \propto \frac{1}{(L/d_i)^3 b_g}. 
\end{equation}
$\gamma$ has an approximately inverse cubic dependence on $L$. In Fig.~\ref{fig:gamma}, for a constant $b_g$, $\gamma/\Omega_{ci}$ drops off very rapidly from the red (higher $\gamma$) to blue (lower $\gamma$) regions as $L$ increases because of this strong dependence on $L$. Contours (dotted lines) denotes $\gamma/\Omega_{ci}$ at 3, 1 (yellow), 0.3 and 0.1 (blue). Simulations having lower $\gamma/\Omega_{ci}$ (of order 0.1) are represented by dots. Simulations having higher $\gamma/\Omega_{ci}$ (of order unity) are represented by diamonds. Parameters and results from four simulations are summarized in Table \ref{tab:param_Vs}. $V_{Ag}$ is based on the initial uniform guide field $B_g$ and the density at the central CS, $n(z_0)=n_0$; note that the modification of $V_{Ag}$ due to temperature anisotropy is insignificant in all simulations because of the strong guide fields. $V_{s\pm}$ is given as the average of the measured spreading speeds from $B_z$ and $U_{ix}$. The ambient ion/electron drift speeds, $U_i$ and $U_e$, are measured at the ambient current sheet in the immediate vicinity of the expanding localized reconnection region (see also \S\ref{bg2-ue} for the determination of ambient $U_e$ and $U_i$ in run {\it Ibg2}). The velocities at the center of the reconnection region, $U_i^{rec}$ and $U_e^{rec}$, are also given for comparison. Note that $U_s^{rec}$ is largely inconsistent with the spreading speeds $V_{s\pm}$ in our simulations, which is consistent with previous zero guide field study \citep{Lapenta:06}. The last column of Table \ref{tab:param_Vs} gives $\theta-\theta_{bi}$, the angle difference between the X-line orientation (of which a detailed study for {\it Ibg1} can be found in \cite{yhliu18b}) and the orientation that bisects the total magnetic shear angle \citep{hesse13a}.



\begin{table}
\begin{center}
\def~{\hphantom{0}}
 \caption{List of simulation parameters and summary of results}\label{tab:param_Vs}
\begin{tabular}{lccccccccccc}
\hline
     & $\gamma/\Omega_{ci}$ & $L/d_i$  & $b_g$  & $V_{Ag}$  & $V_{s+}$ &$V_{s-}$ & $U_i$ & $U_e$ & $U_i^{rec}$ & $U_e^{rec}$& $\theta-\theta_{bi}$  \\[5pt]
\hline
 $\gamma/\Omega_{ci}\sim\mathcal{O}(0.1)$   &&&&&&&&&&& \\
\hline
 {\it Ibg2}    & 0.1  &  0.8   &   2  & $\pm$2 &   0.7 &  -1.0 & 0.7 &  -1 & <0.2 &  -3 &-4$^\circ$\\
 {\it Ibg1}   & 0.3  &  0.8   &   1  & $\pm$1  &  0.7 &  -1.1 & 0.6 &  -1 & <0.1 & -3 & 0$^\circ$\\
\hline
 $\gamma/\Omega_{ci}\sim\mathcal{O}(1)$ &&&&&&&&&&& \\
\hline
 {\it Sbg4}   &  0.6 & 0.4  &  4  & $\pm$4 &  4.1 &  -4.2 & 1.5 &  -1 & 0.6 &  -3.5 &-7$^\circ$\\
 {\it Sbg2}   &  1.2 & 0.4  &  2  & $\pm$2 &   1.7  &   -1.8  & 1.2  &  -1.1 & 0.7 &  -1.5 & -8$^\circ$\\
\hline       
  \end{tabular}
   \end{center}
\end{table}

\subsection{Organization of X-line spreading speeds by $\gamma$}\label{sec:gamma-org}

The distribution of X-line spreading speeds in our simulations can be organized by the ratio of the collisionless tearing growth rate to the ion gyro-frequency, $\gamma/\Omega_{ci}$. Particulary, the sub-Alfv\'enic X-line spreading case ({\it Ibg2}) corresponds to $\gamma/\Omega_{ci}\sim\mathcal{O}(0.1)$ while the Alfv\'enic X-line spreading cases ({\it Sbg4} and {\it Sbg2}) correspond to $\gamma/\Omega_{ci}\sim\mathcal{O}(1)$.

Results from the {\it Ibg2} simulation (upper dot) are presented and discussed in \S\ref{sec:bg2}. The X-line spreads at sub-Alfv\'enic speeds, which are twice slower than $V_{Ag}$ in the system. The X-line spreading speeds in the {\it Ibg1} simulation (bottom dot), which also corresponds to $\gamma/\Omega_{ci}\sim\mathcal{O}(0.1)$, are nearly the same as in run {\it Ibg2}. This suggests that the spreading speed does not directly depend on the guide field strength that is the only key difference between the two runs. In run {\it Ibg1}, however, the guide field $b_g$ = 1, and therefore the ion/electron drift speeds $U_s$ and $V_{Ag}$=$\pm V_{A0}$ are close, making both a possible mediator for the X-line spreading.

Simulations with a sub-ion-scale CS and higher $\gamma/\Omega_{ci}\sim\mathcal{O}(1)$, represented by diamonds, demonstrate Alfv\'enic X-line spreading at $V_{Ag}$, which is significantly higher than the ambient ion/electron drift speeds $U_s$. Results from the {\it Sbg4} simulation (top diamond) are presented in \S\ref{sec:bg4}. The other higher $\gamma$ simulation, {\it Sbg2} (bottom diamond in Fig.~\ref{fig:gamma}), also demonstrates Alfv\'enic X-line spreading at $V_{Ag}$=$\pm$2$V_{A0}$ in that system. Comparison of these two runs shows that when the system has a sufficiently short onset time scale ({\it i.e.}, sufficiently tearing unstable), the Alfv\'enic signal can mediate reconnection onset and lead to X-line spreading. As a result, the X-line spreading speed has a direct and almost linear dependence on the guide field strength \citep{Shepherd:12}.

Our simulations indicate that $\gamma/\Omega_{ci}\gtrsim\mathcal{O}(1)$ is sufficiently strong for the X-line to spread at the Alfv\'en speed $V_{Ag}$\footnote{This ordering may be analogous to a condition on the maximum growth rate of the plasmoid instability proposed for current sheet disruption, i.e., reconnection onset, in resistive magnetohydrodynamic theory \citep{Pucci:14,Uzdensky:16,Huang:17} with the typical reconnection time scale being the Alfv\'en trasit time across the length of the CS and the resulted ordering being $\gamma_{max}\tau_A$ $\gtrsim$ $\mathcal{O}(1)$. Note that in a kinetic plasma, the characteristic time scale based on a typical length scale of $d_i$ and the reconnection Alfv\'en speed $V_{A0}$ is given by $\Omega_{ci}^{-1}=d_i/V_{A0}$.}


\section{Discussions}\label{sec:diss}
\subsection{Comparison with previous numerical studies}\label{sec:diss_1}

Our results compare favorably with a number of previous numerical studies of 3D X-line spreading, in which the direction and speed of X-line spreading are consistent with the current carriers \citep{shay03a,Hesse:05,Lapenta:06,TKMNakamura12,Shepherd:12}. The {\it Ibg2} simulation, having a thicker, ion-scale CS demonstrate X-line spreading at the ion/electron drift speeds, as is well established in zero-guide field simulations \citep{TKMNakamura12}. The thinner, sub-ion-scale CS simulations ({\it Sbg4} and {\it Sbg2}) with faster reconnection onset demonstrate X-line spreading at the guide-field Alfv\'en speed, in agreement with two-fluid simulations with strong guide fields in which a relatively thin equilibrium CS, comparable to or less than the ion sound Larmor radius, is initiated \citep{Shepherd:12}. Both cases lead to a faster onset and allow Alfv\'enic X-line spreading at $V_{Ag}$.

Is the mechanism of X-line spreading mediated by the current carriers (ions/electrons) in a thicker, ion-scale CS case different from that mediated by an Alfv\'enic signal in a thinner, sub-ion-scale CS case? \cite{TKMNakamura12} proposed a picture to explain spreading by the current carriers. Based on pressure balance argument, a pressure decompression region often exists to accompany the initial localized X-line, which is observed in their simulations. The pressure decompression region is then convected by the ion/electron current flow in the out-of-plane direction at $U_i/U_e$, inducing reconnection inflow along $z$ and leading to X-line spreading in the out-of-plane direction. In our sub-Alfv\'enic spreading case (\S\ref{sec:bg2}), we also observe pressure decompression spreading at comparable speeds to the measured $V_{s\pm}$. While the picture invokes reconnection onset, the microphysics of how this happens needs to be elucidated in future investigation. 

The importance of reconnection onset on X-line spreading leads to an interesting question of whether the X-line could spread at all if the equilibrium CS is much thicker than (sub)ion-scale current sheets often initiated in simulations. Indeed, in a PIC simulation with an initial CS thickness $\gtrsim d_i$, it was found that the spreading of the X-line slows down by approximately an order of magnitude compared to a $< d_i$-thick CS simulation \citep{Lapenta:06}. When the CS thickness is twice $d_i$, the X-line even appears to be simply drifting, without clearly extending \citep{shay03a}. Recent work \citep{YHLiu:19a} indicates that a localized X-line can be well confined between a thick ambient CS of 8 $d_i$. The above mentioned studies use a reduced mass ratio of $<$100. In realistic systems, given the wider scale of separation, a non-spreading state of the X-line, suggested in simulations, may correspond to even thicker current sheets. More investigation is required to understand X-line spreading in thick CS systems relevant to space plasmas.

\subsection{Transition between sub-Alfv\'enic and Alfv\'enic X-line spreading}\label{sec:diss_2}

Using the collisionless tearing growth rate $\gamma$, we are able to estimate the onset time scale of a CS. 
Our simulations indicate that $\gamma/\Omega_{ci}\gtrsim\mathcal{O}(1)$ is sufficiently strong for Alfv\'enic X-line spreading. This corresponds to a sufficiently short onset time scale. In contrast, $\gamma/\Omega_{ci}\sim\mathcal{O}(0.1)$ appears to be insufficient for Alfv\'enic X-line spreading, and the X-line spreads at below $V_{Ag}$. The value of $\gamma/\Omega_{ci}$ that marks the transition from sub-Alfv\'enic to Alfv\'enic X-line spreading is expected to lie between $\sim\mathcal{O}(0.1)$ and $\sim\mathcal{O}(1)$. Where this transition occurs and why deserve future investigation, potentially using larger scale numerical simulations.


\section{Implications and Conclusions}\label{sec:imp}

\begin{figure}[htb]
\centering
\hbox{ \hfill \resizebox{3.in}{!}{\includegraphics[scale=1.,trim=0.5cm 7.cm 1cm 9cm, clip=true]{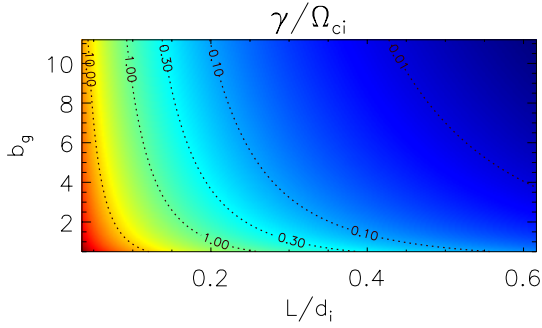}} \hfill 
}
\caption{ \label{fig:gamma-real} Calculation of the normalized collisionless tearing growth rate as a function of CS thickness and guide-field strength $\gamma (L/d_i,b_g)/\Omega_{ci}$ in logarithmic scale using parameters from the magnetopause X-line spreading event. Same format as Fig.~\ref{fig:gamma}. } 
\end{figure}

{\it Magnetopause}: the CS (half) thickness is typically $L\sim$ 5 $d_i$ based on statistical multispacecraft measurements \citep{BerchemRussell:82}. The corresponding $\gamma/\Omega_{ci}$ is unlikely to reach $\sim\mathcal{O}(1)$. Calculation of $\gamma$ similar to Fig.~\ref{fig:gamma} using parameters from the recent magnetopause observation \citep{Zou:2018}, plotted in Fig.~\ref{fig:gamma-real}, reveals that the CS thickness such that $\gamma/\Omega_{ci}\sim$ 1 corresponds to a sub-ion-scale thickness of $L\sim$ 0.2 $d_i$, which would require strong compression from the statistical thickness of $\sim$ 5 $d_i$. While it appears to be stringent, our study suggests that under typical conditions at the magnetopause, the reconnection X-line is unlikely to demonstrate Alfv\'enic spreading at the local Alfv\'en speed regardless of the guide field strength.



{\it Turbulent Magnetosheath}: reconnection of electron-scale current sheets that do not couple to ions has recently been reported \citep{Phan:18}. Well described by 2D reconnection simulations using similar conditions \citep{SharmaPyakurel:2019}, it is consistent with the picture of elongated turbulent magnetic flux tubes naturally developed in anisotropic plasma turbulence \citep{goldreich95a}. Such reconnecting electron-scale current sheets will favor X-line spreading mediated by an Alfv\'enic signal along the elongation direction under a strong guide field, which is present in most sub-ion-scale current sheets observed \citep{Phan:18}. The turbulent magnetosheath thus represents a space system to plausibly realize Alfv\'enic, out-of-plane expansion of localized reconnection X-lines. 

We have investigated the spreading of localized X-lines out of the reconnection plane using large-scale 3D PIC simulations with varying current sheet thicknesses and guide fields. The physics of reconnection onset is found to be important for X-line spreading. In a simulation with an ion-scale equilibrium current sheet, resulting in a slower reconnection onset, the X-line does not spread at the Alfv\'en speed (based on the guide field) $V_{Ag}$ even in the presence of a strong guide field; instead, it spreads at the significantly slower, ion/electron drift speeds, as previously established in zero guide field studies. We further show that the responsible ion/electron drift speeds for spreading are determined at the immediate ambient CS or the leading-edge (rather than at the center) of the expanding reconnection region.

In simulations with a thinner, sub-ion-scale equilibrium current sheet, the X-line spreads at the Alfv\'en speed $V_{Ag}$, which is significantly higher than the ambient ion/electron drift speeds. An Alfv\'enic signal with characteristics consistent with kinetic Alfv\'en waves develops from localized initial perturbations prior to reconnection onset. It propagates bidirectionally, thins the CS and extends the CS thinning at $\sim V_{Ag}$, leading to the spreading of reconnection onset at approximately the same speed ({\it i.e.} Alfv\'enic X-line spreading).


We have demonstrated the strong dependence on the current sheet thickness of the X-line spreading speed. The X-line orientations in our simulations are found to be consistent with the fastest growing modes of the collisionless tearing instability. The tearing growth rate $\gamma$ organizes the X-line spreading speeds into Alfv\'enic and sub-Alfv\'enic. Our simulations indicate a tearing growth rate of $\gamma/\Omega_{ci}\sim\mathcal{O}(1)$ is sufficiently strong for Alfv\'enic X-line spreading, while $\gamma/\Omega_{ci}<\mathcal{O}(1)$ is insufficient and can only realize the slower, sub-Alfv\'enic X-line spreading regardless of strong guide field conditions.

Our results compare favorably with a number of numerical simulations and recent magnetopause observations. They are relevant for the upcoming ESA-CAS joint mission, Solar wind Magnetosphere Ionosphere Link Explorer (SMILE), which will study the development of reconnection-lines at Earth's magnetopause using x-ray and UV imagers. A key implication of this work is that the typically thick magnetopause current sheet must be substantially compressed to a state strongly unstable to collisionless tearing such that Alfv\'enic X-line spreading can effectively take place. Otherwise, an X-line will likely spread at speeds well below the local Alfv\'en speed. 


\section{Appendix: The collisionless tearing growth rate}\label{appx}
The derivation that gives the analytical form of the collisionless tearing growth rate is outlined as follows. Consider the collisionless tearing stability of the magnetic profile used in the simulations for an arbitrary wavevector ${\bf k}=k_x\hat{{\bf x}}+k_y\hat{{\bf y}}$ that makes an oblique angle $\theta \equiv \tan^{-1}(k_y/k_x)$ with $\hat{{\bf y}}$ and resonance surface $z_s$ defined by $F|_{z=z_s} \equiv {\bf k \cdot B}|_{z=z_s}=0$. In the outer region, the magnetohydrodynamic model gives an eigenmode equation \citep{furth63a} of the form $\tilde{\psi}'' =(k^2+F''/F)\tilde{\psi}$, where $\tilde{\psi}(z)$ is the perturbed flux function on the oblique plane and $k^2 \equiv k_x^2 +k_y^2$. We combine the approximate solutions for $kL \ll 1$ and $kL \gg 1$ as in \citep{baalrud12a}, and obtain the drive for tearing perturbations \citep{furth63a} $\Delta'\equiv \lim_{\epsilon\rightarrow 0} (1/\tilde{\psi})[d\tilde\psi/dz]_{z_s-\epsilon}^{z_s+\epsilon}\simeq (\alpha^2/k)(F^{-2}_{-\infty}+F^{-2}_{\infty})-2k$ where $\alpha\equiv(dF/dz)_{z=z_s}$. Substituting our magnetic profile, we get 
\begin{equation}
\Delta' \simeq \frac{2[(1/2+b_g\mbox{tan}\theta)^2+1]}{kL^2}-2k.
\label{dprime_eqn}
\end{equation}
The upper bound of the unstable wavenumber is $k_cL\simeq [(1/2+b_g\mbox{tan}\theta)^2+1]^{1/2}$. The standard matching approach \citep{drake77a,daughton11a} at the kinetic resonance layer gives the growth rate
\begin{equation}
\gamma_t \simeq \frac{d_e^2\Delta'}{l_s}kv_{the},
\label{eq:gamma}
\end{equation}
where $v_{the} \equiv (2 T_e/m_e)^{1/2} $ is the electron thermal speed and $d_e \equiv c/\omega_{pe}$ is the local electron inertial length at the resonant surface. $l_s$ is the scale length of the magnetic shear defined in $k_{\parallel} = {\bf k} \cdot {\bf B}/|{\bf B}|\approx [\partial ({\bf k}\cdot{\bf B}/|{\bf B}|)/\partial z]_{z=z_s} (z-z_s) \equiv k (z-z_s)/l_s$. It is derived to be
 \begin{displaymath}
  l_s= \frac{L b_g (1+\mbox{tan}^2\theta)^{1/2}}{[1-(1/2+b_g\mbox{tan}\theta)^2]\mbox{cos}\theta}.
\end{displaymath} 
The dominant mode typically has a wavelength $kL\sim k_cL/2$ and it is $\sim0.5$. Using the wavelength of the dominant mode, i.e., taking $kL\sim k_cL/2$, we derive $\gamma_t$=$\gamma_t(\theta,L/d_e,b_g,v_{the}/c)$ as a function of the various parameters. The $\gamma$ used in \S\ref{sec:gamma_sim} is the maximum of $\gamma_t$ over all oblique angles $\theta$. For constant $v_{the}$ used in all simulations, $\gamma$ is reduced to a function of the CS thickness and guide field strength, $\gamma(L/d_i,b_g)$.

\acknowledgments
TCL is thankful for J. Drake, M. Shay, N. Loureiro, K.M. Schoeffler, L.-J. Chen, C. Norgren, N. Bessho, S. Wang and K. Knizhnik for invaluable discussions. TCL and YHL are supported by NASA grants 80NSSC18K0754 and MMS mission 80NSSC18K0289. MH acknowledges support by the Research Council of Norway/CoE under contract 223252/F50 and by NASA's MMS mission. YZ is supported by NSF grant AGS-1664885 and Jack Eddy Postdoctoral Fellowship from UCAR's Cooperative Programs for the Advancement of Earth System Science (CPAESS). Simulations are supported by NSF- Blue Waters Petascale Computing Resource Allocation project no.~ACI1640768 and NERSC Advanced Supercomputing. Blue Waters is supported by NSF OCI-0725070 and ACI-1238993 awards, and is a joint effort of UIUC and its NCSA. The large data generated by peta-scale PIC simulations can hardly be made publicly available. Interested researchers are welcome to contact the corresponding author for subset of the data archived in computational centers.


\end{document}